\documentstyle[preprint,aps,eqsecnum]{revtex}
 \tightenlines
 \begin{document}

 \draft

 \title{Bogomolny's semiclassical transfer operator \\
 for rotationally invariant integrable systems}
 \author{D.A.\ Goodings and N.D.\ Whelan}
 \address{Department of Physics and Astronomy, McMaster University,\\
 Hamilton, Ontario, Canada L8S 4M1}
 \date{March 19, 1998}
 \maketitle

 \begin{abstract}
 The transfer operator due to Bogomolny provides a convenient method for
 obtaining a semiclassical approximation to the energy eigenvalues of
 a quantum system, no matter what the nature of the analogous classical
 system. In this paper, the method is applied to integrable systems
 which are rotationally invariant, in two and three dimensions.
 In two dimensions, the transfer operator is expanded in a Fourier
 series in the angle variable, while in three dimensions it is
 expanded in spherical harmonics. In both cases, when the Fourier
 coefficients are evaluated using the stationary phase approximation,
 we arrive at the Einstein-Brillouin-Keller (EBK) quantization conditions.
 The associated Maslov indices are shown to agree with the results
 calculated by well-known simple rules. The theory is applied to several
 rotationally invariant systems, including the hydrogen atom and the
 isotropic harmonic oscillator in two and three dimensions, the circle
 billiard, a billiard inside a spherical cavity, and a harmonic potential
 with a singular magnetic flux line.

 \end{abstract}

 \section{Introduction}
 \label{intro}
 In furthering our understanding of the relationship between classical
 mechanics and quantum mechanics, semiclassical approximations play an
 important role. Periodic orbit theory, developed by Gutzwiller,
 Balian and Bloch, Berry and others \cite{G90,Chaos92,BB97}
 employs the periodic orbits of the
 classical system to obtain a semiclassical approximation to the density
 of states or to individual energy eigenvalues of the analogous
 quantum system. Although formally elegant and satisfying, the theory
 is usually hard to apply in practice, because the periodic orbit sum
 is not absolutely convergent, and because it is difficult to find the
 periodic orbits in a systematic way. Alternative semiclassical
 approximations, which do not depend on knowing the periodic orbits,
 have been proposed by Bogomolny \cite{Bog92,Bog92a} and by Doron and
 Smilansky \cite{DS92,DS93}. Exploiting the duality between the classical
 dynamics of a billiard inside a bounded region and the scattering of
 external particles by the system's boundary, Doron and Smilansky obtained
 semiclassical energy eigenvalues for the billiard system by
 constructing a semiclassical approximation to the scattering matrix
 for the exterior problem. In the theory proposed by Bogomolny, one
 chooses a Poincar\'e surface of section (PSS) which is frequently
 crossed during the motion of the system, and one constructs a
 semiclassical transfer operator from the classical trajectories which
 take the system from one position on the PSS to another. For billiard
 systems, the approaches based on the scattering matrix and on the transfer
 operator can be shown to yield the same determinantal equation for the
 energy eigenvalues of the interior system \cite{Bog92,RS95}.

 The present paper had its origin in trying to derive the correct
 quantum energy eigenvalues of the hydrogen atom (in three dimensions)
 by means of Bogomolny's semiclassical transfer operator. By adding a
 small $1/r^2$ term to the Coulomb potential---a device which prevents
 the transfer operator from being singular---we achieved this goal. However,
 an essential part of the derivation was the use of the stationary
 phase approximation. Subsequently, we generalized our approach
 and showed that it led to the well-known Einstein-Brillouin-Keller
 (EBK) quantization rules \cite{Einstein17,Brillouin26,Keller58,KR60}.
 When viewed in this light, there is nothing special about our solution
 for the hydrogen atom. In fact, the main result of our paper can be
 summed up concisely: for rotationally invariant integrable systems,
 Bogomolny's transfer operator, plus the stationary phase approximation,
 yields EBK quantization. In a separate publication \cite{WG98} we plan
 to show how the EBK quantization rules can also be obtained from Bogomolny's
 transfer operator formulated in terms of the angle-action variables.

 The plan of the paper is as follows. After a brief description of the
 transfer operator in the next section, in Sec.\ \ref{sec3} we construct the
 transfer operator for a two-dimensional system having circular symmetry.
 By making a suitable Fourier expansion and evaluating the Fourier
 coefficients by means of the stationary phase approximation, we derive
 the EBK quantization conditions. This general formulation is applied
 in Sec.\ \ref{sec4} to the hydrogen atom (plus $1/r^2$ potential), the
 circular harmonic oscillator (plus $1/r^2$ potential), a harmonic
 potential plus a singular magnetic flux line, the circle billiard, and
 the annulus billiard. In Sec.\ \ref{sec5} a similar approach is described
 for three-dimensional systems having spherical symmetry. The resulting
 EBK quantization conditions are applied in Sec.\ \ref{sec6} to the hydrogen
 atom (plus $1/r^2$ potential), the isotropic harmonic oscillator (plus
 $1/r^2$ potential), and a billiard inside a spherical cavity. The
 paper concludes with a discussion of our results for the EBK energy
 eigenvalues in comparison with the exact quantum energies.

 \section{The transfer operator and the determinantal equation}
 \label{sec2}
 We begin with a brief description of Bogomolny's semiclassical transfer
 operator. For a system with $f$ freedoms, the PSS in configuration
 space is a surface or hypersurface of dimension $f-1$, and the transfer
 operator in the coordinate representation is \cite{Bog92}
 \begin{equation}
 T(q^{\prime\prime},q^\prime;E)=\sum\limits_{cl.tr.}{1\over{(2\pi i\hbar)
 ^{(f-1)/2}}}\left|{\rm det}{{\partial^2 S(q^{\prime\prime},q^\prime;E)}\over
 {\partial q^{\prime\prime} \partial q^\prime}} \right|^{\frac{1}{2}}
 \exp[iS(q^{\prime\prime},q^\prime;E)/\hbar-i\mu\pi/2],
 \label{eq1}
 \end{equation}
 where $q^\prime$ and $q^{\prime\prime}$ denote $f-1$ generalized
 coordinates for two points located on the PSS.
 The summation is over all classical trajectories which go from $q^\prime$
 to $q^{\prime\prime}$, crossing the PSS at these points in the same sense
 and at no other points (in the same sense) in going from $q^\prime$ to
 $q^{\prime\prime}$.
 For each such trajectory one needs the action at energy $E$, denoted by
 $S(q^{\prime\prime},q^\prime;E)$, and the phase index $\mu$, which is
 related to the occurrence of caustics---points on the trajectory at which
 the semiclassical approximation is not valid. The matrix of second
 derivatives of the action has dimension $f-1$.
 
 If $T(E)$ is the transfer operator and $I$ is the unit operator, the
 corresponding semiclassical energy eigenvalues of the quantum
 system are determined from the condition \cite{Bog92}
 \begin{equation}
 {\rm det}[I-T(E)]=0.
 \label{eq2}
 \end{equation}
 When properly formulated, the $T$-operator is unitary \cite{Bog92}.
 In the past few years there have
 been several applications of the transfer operator based on
 constructing an approximation to $T(E)$ in coordinate space
 \cite{Lauritzen92,SLG93,SLG94,Boas94,Hag95,Lefebvre95,B96,B96a,BGL96,TG97}.
 However, when the system being treated has rotational symmetry, it
 is better to treat the transfer operator in the angular momentum
 representation, since it is then diagonal
 \cite{Lefebvre95}. In the next section we show how this may be
 carried out for two-dimensional systems with circular symmetry.

 \section{Two-dimensional systems with circular symmetry}
 \label{sec3}
 Let us consider a particle of unit mass moving in two dimensions in
 a potential $V(r)$. The Hamiltonian is
 \begin{equation}
 H=\frac{p_r^2}{2}+\frac{p_\phi^2}{2r^2}+V(r)
 \label{eq3}
 \end{equation}
 where $p_r=\dot r$ and $p_\phi=r^2\dot\phi$ are the momenta conjugate
 to the polar coordinates $r$ and $\phi$ describing the particle's
 position. Since the angular momentum $p_\phi$ is a constant of the
 motion, let us denote it as $L$. For given $E$ and $L$, the turning points
 of the classical motion along the radial direction are determined by
 \begin{equation}
 \frac{L^2}{2r^2} + V(r) = E.
 \label{eq4}
 \end{equation}
 Clearly, the turning point radii, $r_-$ and $r_+$, depend on $L$ as
 well as $E$.

 In setting up the transfer operator, we choose the Poincar\'e surface
 of section (PSS) to be a circle of radius $R$. While in principle any
 radius between $r_-$ and $r_+$ could be used, there is a natural
 choice for given energy $E$. At that energy there is a trajectory
 with maximal angular momentum $L_{max}(E)$ which is a circle, corresponding
 to the radial kinetic energy being zero. We define $R$ to be the radius of
 this circle. It can also be thought of as the circle for which the
 radial turning points coincide. Then, trajectories at energy $E$
 having $|L|<|L_{max}(E)|$ have nonzero radial kinetic energy, and,
 therefore, must repeatedly cross this circle, making it a suitable
 choice for the PSS. (It is easy to show from equation (\ref{eq4})
 and the derivative of this equation that $R$ is the solution of
 $2V(r)+rV^\prime(r)=2E$.)

 The coordinate $q$ on the PSS will be taken to be the polar coordinate
 $\phi$. From equation (\ref{eq1}) the transfer operator from $\phi$ to
 $\phi^\prime$ (where $0\le\phi\le 2\pi$ and $0\le\phi^\prime\le 2\pi$)
 on the PSS is
 \begin{equation}
 T(\phi^\prime,\phi;E)=\sum_j{1\over{(2\pi i\hbar)^{1/2}}}
 \left|{{\partial^2 S_j(\phi^\prime,\phi;E)}\over
 {\partial\phi^\prime \partial\phi}} \right|^{1/2}
 \exp[iS_j(\phi^\prime,\phi;E)/\hbar-i\mu_j\pi/2],
 \label{eq5}
 \end{equation}
 where $j$ labels different possible classical trajectories at energy
 $E$ which go from $\phi$ to $\phi^\prime$ without crossing
 the PSS (in the same sense) at any other point.
 Let us define the angle $\gamma$ to be $\gamma=\phi^\prime-\phi$
 (modulo $2\pi$) in order that $0\le\gamma\le 2\pi$. The trajectories
 can be labelled in such a way that the angle traversed by the particle
 in going from $\phi$ to $\phi^\prime$ is $\xi^{(j)}= \gamma+2\pi j$,
 where $0\le\gamma\le 2\pi$ and $j$ is an integer
 (positive, negative or zero). The action along the $j$th trajectory
 can be written as $S(\xi^{(j)};E)$. Also, we can write
 \begin{equation}
 \displaystyle \frac{\partial^2S(\xi^{(j)};E)}{\partial\phi\partial
 \phi^\prime} = -\frac{\partial^2S(\xi^{(j)};E)}{\partial\gamma^2}.
 \label{eq6}
 \end{equation}
 Thus, in accord with the invariance of the Hamiltonian under
 rotations about the origin, the transfer operator can be expressed in
 terms of the relative angle $\gamma$ only. We can, therefore, expand
 (\ref{eq5}) in a Fourier series:
 \begin{equation}
 T(\phi^\prime,\phi;E)=T(\gamma;E)
 = \displaystyle \sum_{m=-\infty}^\infty C_m(E)\exp(im\gamma),
 \label{eq7}
 \end{equation}
 where the expansion coefficients are
 \begin{equation}
 C_m(E) = \displaystyle \frac{1}{2\pi} \int_0^{2\pi} T(\gamma;E)
 \exp(-im\gamma) d\gamma.
 \label{eq8}
 \end{equation}

 We now construct a matrix representation of the transfer operator
 using the basis $\{(2\pi)^{-1/2} \exp(im\phi)\}$. Since
 $\gamma=\phi^\prime-\phi$ (modulo $2\pi$), a typical matrix element is
 \begin{equation}
 T_{m_1m_2}(E)=\displaystyle \frac{1}{2\pi} \int_0^{2\pi} d\phi
 \int_0^{2\pi} d\phi^\prime \exp(-im_1\phi^\prime) T(\gamma;E)
 \exp(im_2\phi) = 2\pi\, C_{m_1}(E) \delta_{m_1m_2}.
 \label{eq9}
 \end{equation}
 Thus, the $T$-matrix is diagonal in this representation, and its
 eigenvalues (as a function of $E$) are just the diagonal elements.
 (Note that these $T$-matrix eigenvalues should not be confused
 with the semiclassical energy eigenvalues.) Denoting
 the $m$th eigenvalue curve as $\lambda_m(E)$, we obtain from equations
 (\ref{eq5}) to (\ref{eq9}),
 \begin{eqnarray}
 &\displaystyle\lambda_m(E)=\int_0^{2\pi} T(\gamma;E) \exp(-im\gamma)
 d\gamma \nonumber\\
 &\displaystyle ={1\over{(2\pi i\hbar)^{1/2}}} \sum_j \exp(-i\mu_j\pi/2)
 \int_0^{2\pi} d\gamma \left|{{\partial^2 S(\xi^{(j)};E)}\over
 {\partial\gamma^2}} \right|^{1/2}
 \exp[iS(\xi^{(j)};E)/\hbar-im\gamma].
 \label{eq10}
 \end{eqnarray}
 We remind the reader that $\xi^{(j)}$ depends on $\gamma$ through the
 definition $\xi^{(j)}=\gamma+2\pi j$.

 Up to this point we have made no approximations other than the
 approximation involved in deriving Bogomolny's semiclassical transfer
 operator. We now evaluate the integrals in (\ref{eq10}) using the
 stationary phase approximation. For the $j$th integral the point at
 which the phase is stationary is determined by the equation
 \begin{equation}
 \displaystyle \frac{\partial S(\xi^{(j)};E)}{\partial\gamma} = m\hbar.
 \label{eq11}
 \end{equation}
 The left-hand side of this relation is the classical angular momentum
 $L$ for the $j$th trajectory. Thus, the stationary phase condition
 effectively quantizes the angular momentum of the particle.
 Because the possible trajectories from $\phi$ to $\phi^\prime$
 at energy $E$ are uniquely specified by the angular momentum, it is
 clear that for each value of $m$ (positive, negative or zero) there
 is {\it at most one trajectory} satisfying (\ref{eq11}). (A solution
 exists if and only if $|L|\le |L_{max}(E)|$.) The value
 of $j$ for this trajectory will be denoted $j_m$. Thus, for given $m$,
 we denote the solution of equation (\ref{eq11}) (when it exists) as
 $\gamma_m$, and the corresponding angle traversed by the particle in
 going from $\phi$ to $\phi^\prime$ as $\xi_m=\gamma_m+2\pi j_m$.

 When a solution $\gamma_m$ of equation (\ref{eq11}) exists for a
 particular value of $j_m$, we can evaluate the integral in the usual
 way, assuming that $\partial^2 S/\partial\gamma^2$
 is a relatively slowly varying function of $\gamma$. Introducing the
 symbol $\nu_m$ through the definition
 \begin{eqnarray}
 & \nu_m=0 \qquad{\rm if}\qquad \displaystyle \left( \frac
 {\partial^2S}{\partial\gamma^2} \right)_{\gamma=\gamma_m}>0 \nonumber\\
 & \nu_m=1 \qquad{\rm if}\qquad \displaystyle \left( \frac
 {\partial^2S}{\partial\gamma^2} \right)_{\gamma=\gamma_m}<0
 \label{eq12}
 \end{eqnarray}
 and henceforth denoting $\mu_j$ for the trajectory $j_m$ as $\mu_m$,
 we obtain
 \begin{equation}
 \lambda_m(E)\approx\exp[iS(\xi_m;E)/\hbar-im\gamma_m
 -i(\mu_m+\nu_m)\pi/2].
 \label{eq13}
 \end{equation}
 The fact that these approximate eigenvalues of the $T$-matrix have
 unit modulus is consistent with the $T$-matrix being unitary
 \cite{Bog92}. Note that the symbols $\mu_m$ and $\nu_m$ have the same
 meaning as in the paper by Creagh, Robbins and Littlejohn \cite{CRL90}.

 It is useful to further simplify this expression by splitting the
 action for the trajectory $j_m$ into radial and angular parts. The
 angular part, evaluated at $\gamma_m$, may be denoted by
 \begin{equation}
 S_{\rm ang}(\xi_m;E)=\int p_\phi d\phi = m\hbar(\gamma_m+2\pi j_m),
 \label{eq14}
 \end{equation}
 and the radial part, evaluated at the angular momentum $L=m\hbar$
 determined by the stationary phase condition, is
 \begin{equation}
 S_{\rm rad}(L=m\hbar;E)=\oint |p_r|\,|dr| = 2\int_{r_-}^{r_+} |p_r|\,dr.
 \label{eq15}
 \end{equation}
 Using the fact that $\exp(im2\pi j_m)=1$, we arrive at the expression
 \begin{equation}
 \lambda_m(E)\approx\exp[iS_{\rm rad}(L=m\hbar;E)/\hbar-i(\mu_m+\nu_m)\pi/2].
 \label{eq16}
 \end{equation}
 
 The semiclassical energy eigenvalues of the quantum system are found
 from the determinantal equation (\ref{eq2}), which is satisfied
 whenever an eigenvalue of the $T$-matrix is equal to unity. Thus, the
 condition for an energy eigenvalue is that
 $\lambda_m(E)=\exp(i2\pi n_r)$. From equation (\ref{eq16}) this yields
 \begin{equation}
 S_{\rm rad}(L=m\hbar;E)=2\pi\hbar(n_r+\sigma_m/4)\qquad\qquad n_r=0,1,2,\dots
 \label{eq16a}
 \end{equation}
 where we define the Maslov index $\sigma_m=\mu_m+\nu_m$. This is
 associated with a complete cycle of the radial motion for a trajectory
 at energy $E$ and angular momentum $L=m\hbar$. The allowed values of
 $n_r$ in (\ref{eq16a}) are determined by the assumption that
 $S_{\rm rad}\ge 0$.

 In Appendix \ref{appA} it is shown that, for smooth
 potentials, the combination $\sigma_m=\mu_m+\nu_m$ is always equal to 2. 
 This is the result one would obtain for the Maslov index in EBK
 quantization \cite{Keller58,KR60}
 using the simple rule of counting 1 for each of the soft turnarounds
 during a complete cycle of the radial motion. It is also shown in
 Appendix \ref{appA} that if the particle is confined inside a circular
 disk with a hard wall (Dirichlet boundary condition on the wave function),
 the result for $\sigma_m$ is 3. This agrees with the simple
 rule for computing the EBK Maslov index by counting 1 for the soft
 turnaround at the inner radial turning point and 2 for the collision
 with the disk boundary. Thus, for systems having circular symmetry
 in two dimensions, Bogomolny's transfer operator (modified using
 the stationary phase approximation) leads to EBK
 quantization, with the Maslov index for the radial motion computed by
 the well-known simple rules. Note that $\sigma_m$ is a canonical
 invariant, even though the transfer operator is not canonically
 invariant (since it depends on the choice of the PSS).
 In addition to (\ref{eq16a}), the other
 EBK quantization condition, $L=m\hbar$, was, of course, obtained from
 the stationary phase condition (\ref{eq11}). It is worth noting that
 for systems that are invariant under time reversal, the energy
 eigenvalues with $m\ne 0$ are doubly degenerate.

 \section{Application to systems in two dimensions}
 \label{sec4}
 \subsection{The Coulomb plus $1/r^2$ potential}
 \label{sec4A}
 Let us now apply this general formulation to the hydrogen atom in two
 dimensions. In fact we shall treat a slightly more complicated potential,
 namely the Coulomb potential plus a term proportional to $1/r^2$. For
 the pure Coulomb potential, there is a one-parameter family of ellipses
 which start out from a given point on the PSS and return to the same
 point. Thus, this point is a focal point. By adding the $1/r^2$ term,
 we ensure that the trajectories are not ellipses, and thereby avoid
 difficulties associated with the initial point being a conjugate point.

 Assuming that the nucleus is stationary at the origin, we take the
 potential to be
 \begin{equation}
 \displaystyle V(r)=-\frac{1}{r} \pm \frac{\alpha^2}{2r^2}.
 \label{eq17}
 \end{equation}
 For convenience we have taken the electronic charge $e$ to be unity, and
 we have written the strength of the $1/r^2$ potential as $\alpha^2/2$,
 where $\alpha$ has the dimensions of angular momentum.
 The $1/r^2$ potential may be repulsive or attractive, and provided
 $\alpha^2$ is not too large in the repulsive case,
 the electron will always be bound to the nucleus, implying that the
 energy $E$ is negative. From equation (\ref{eq4}), the classical turning
 points of the radial motion occur at radii $r_-$ and $r_+$ given by
 \begin{equation}
 r_{\pm}=\displaystyle \frac{1\pm\beta}{2|E|}
 \label{eq18}
 \end{equation}
 \begin{equation}
 \beta=[1-2|E|(L^2\pm\alpha^2)]^{1/2}.
 \label{eq19}
 \end{equation}
 The radius of the Poincar\'e circle defined in the previous section is
 $R=1/(2|E|)$ (although we shall not make explicit use of this in what
 follows). For the repulsive $1/r^2$ potential, each trajectory
 traverses an angle less than $2\pi$ before returning to the PSS, while
 in the attractive case, the trajectories go through angles greater than
 $2\pi$. This means that the possible trajectories are qualitatively
 different in the two cases, there being only two possibilities in the
 former case and many in the latter case.
 Note from (\ref{eq19}) that, for given $E$, the maximum
 possible value of the classical angular momentum is given by
 \begin{equation}
 [L_{max}(E)]^2= \frac{1}{2|E|} \mp\alpha^2.
 \label{eq20}
 \end{equation}

 To obtain the semiclassical energy eigenvalues from equation
 (\ref{eq16a}), we must calculate the radial action integral with
 $L=m\hbar$, as in (\ref{eq15}). Using equation (\ref{eq3}) with
 $L=m\hbar$ to solve for $|p_r|$ as a function of $r$, we obtain
 \begin{equation}
 S_{\rm rad}(L=m\hbar;E) =2\int\limits_{r_-}^{r_+}|p_r|dr
 =\pi\left( \frac{2}{|E|} \right)^{1/2} -2\pi(m^2\hbar^2\pm\alpha^2)^{1/2}
 \label{eq21}
 \end{equation}
 Here the first term on the right-hand side is the action of each member
 of the family of elliptical orbits at energy $E$ of the pure Coulomb
 potential. To write down the EBK quantization condition from equation
 (\ref{eq16a}), we set $\sigma_m=2$, corresponding to two soft turnarounds
 at the radial turning points (see Appendix \ref{appA}). Hence,
 \begin{equation}
 \displaystyle \frac{\pi}{\hbar}\left(\frac{2}{|E|}\right)^{1/2}
 -2\pi(m^2\pm\alpha^2/\hbar^2)^{1/2}=(2n_r+1)\pi, \qquad\qquad
 n_r=0,1,2,\dots
 \label{eq24}
 \end{equation}
 Using the fact that $E$ is negative for the bound state solutions we
 are considering, we obtain
 \begin{eqnarray}
 &\displaystyle E_{mn_r}=-\frac{1}{2\hbar^2[n_r+\frac{1}{2}
 +(m^2\pm\alpha^2/\hbar^2)^{1/2}]^2}, \nonumber\\
 &\qquad m=0,\pm 1,\pm 2,\dots \qquad n_r=0,1,2,\dots
 \label{eq25}
 \end{eqnarray}
 This expression gives the approximate semiclassical energy
 eigenvalues for the Coulomb plus $1/r^2$ potential. Note that the
 energy is the same for positive and negative values of $m$,
 implying that the energy eigenvalues are doubly degenerate for $m\ne 0$
 and nondegenerate for $m=0$. The allowed values of $m$ are constrained
 by the condition $|m|\hbar\le |L_{max}(E)|$, with $|L_{max}(E)|$ given
 by equation (\ref{eq20}).
 
 The pure Coulomb potential is obtained by letting $\alpha\to 0$ in
 equation (\ref{eq25}). Until now we have assumed that $\alpha$ is
 nonzero and sufficiently large (presumably, $\alpha \gg \hbar$) that
 the trajectories are not close to being ellipses, thereby avoiding
 the point $\phi^\prime$ on the PSS being a focal point. At this
 stage, however, it is permissible to relax this requirement and let
 $\alpha$ become zero. Putting $n=|m|+n_r$, we obtain
 \begin{equation}
 E_n=-\frac{1}{2\hbar^2(n+\frac{1}{2})^2}, \qquad\qquad n=0,1,2,\dots
 \label{eq26}
 \end{equation}
 (It is satisfying that the limit does not depend on the sign of the
 $\alpha^2/(2r^2)$ term in the potential, despite the fact that the
 two problems are quite different, as mentioned earlier.) 
 This expression for the energy eigenvalues is exactly the same as the
 result found by solving the two-dimensional Schr\"odinger equation
 for the Coulomb potential. The eigenvalues do not depend explicitly
 on $m$, but it is clear from the definition of $n$ that $|m|\le n$.
 This condition, which also arises, for example, in solving the radial
 Schr\"odinger equation by the method of series expansion, or by using
 group-theoretical considerations, correctly
 determines the degeneracies of the energy levels given by (\ref{eq26}).

 In an earlier study of the hydrogen atom in two dimensions using
 Bogomolny's transfer operator \cite{BGL96}, the PSS was chosen to be
 a radial line. The ``half-mapping'' transfer operators introduced by
 Haggerty \cite{Hag95} were used to avoid the problem associated with
 the family of trajectories (ellipses) at energy $E$ starting from a point
 on the PSS and returning to the same point on the PSS. The outcome of
 this work was similar to equation (\ref{eq26}) but with $(n+{1\over 2})^2$
 replaced by $(n+{3\over 4})^2$. We now realize that this peculiar
 result was due to an incorrect assignment of the phase indices
 associated with the elliptical trajectories. In Fig.\ 2 of Ref.\ 
 \cite{BGL96}, there is a caustic associated with the longer solid
 trajectory, but there is {\it no} caustic associated with the shorter
 solid trajectory. When this fact is properly taken into account, the
 energy levels turn out to be the same as equation (\ref{eq26}). Thus,
 the energy levels of this system do not depend on the
 choice of the PSS, at least for the two choices considered.

\subsection{The circular harmonic oscillator plus $1/r^2$ potential}
\label{sec4B}
A particle moving in two dimensions in a circular harmonic oscillator
plus $1/r^2$ potential has been treated in an earlier paper \cite{TG97}
using a slightly different method based on Bogomolny's transfer operator.
Here we show that our general formulation of Sec.\ \ref{sec3} leads
quickly to the same results for the energy eigenvalues of this system.

We take the potential to be
\begin{equation}
V(r)=\displaystyle \frac{1}{2} \omega^2 r^2 \pm \frac{\alpha^2}{2r^2},
\label{eq27}
\end{equation}
where, as in the last subsection, $\alpha^2/2$ is the strength of the
$1/r^2$ potential, which may be attractive or repulsive. When the
particle has energy $E$, equation (\ref{eq4}) leads to the following
expression for the classical turning-point radii:
\begin{equation}
r_{\pm}^2=\displaystyle \frac{E\pm[E^2-\omega^2(L^2\pm\alpha^2)]^{1/2}}
{\omega^2}.
\label{eq28}
\end{equation}
Using this, and solving equation (\ref{eq3}) to find $|p_r|$ as a
function of $r$, one finds,
\begin{equation}
S_{\rm rad}(L;E)=2\int_{r_-}^{r_+}|p_r|dr
=\displaystyle \frac{\pi E}{\omega} -\pi(L^2 \pm \alpha^2)^{1/2}.
\label{eq29}
\end{equation}
When this is substituted in equation (\ref{eq16a}), with $L$ set equal
to $m\hbar$ and $\sigma_m$ set equal to 2 (corresponding to soft
turnarounds at $r_-$ and $r_+$), we obtain for the energy eigenvalues
belonging to a given value of $m$
\begin{eqnarray}
&E_{mn_r}=\hbar\omega[2n_r + (m^2\pm\alpha^2/\hbar^2)^{1/2} +1],\nonumber\\
&m=0,\pm 1,\pm 2,\dots \qquad n_r=0,1,2\dots
\label{eq30}
\end{eqnarray}
This expression agrees with the result obtained from an exact solution
of the Schr\"odinger equation (see, for example, Fl\"ugge \cite{Flugge74}).
A plausible explanation of this agreement, despite the approximations
inherent in our semiclassical approach, has been put forward earlier
\cite{TG97} and will be briefly discussed in the final section of the paper.
Note that the energy eigenvalues are doubly degenerate when $m\ne 0$
and nondegenerate when $m=0$.

\subsection{Circular harmonic oscillator plus singular magnetic flux line}
\label{sec4C}
The development in the last subsection can be extended to include a
singular magnetic flux line passing through the origin. This means
that, in addition to the circular harmonic oscillator potential and
the $1/r^2$ potential, the particle motion occurs in the presence of a
magnetic field (perpendicular to the plane of the motion) having the
form of a $\delta$-function singularity at the origin. The magnetic
field breaks the time-reversal symmetry and removes the degeneracy of
the energy eigenvalues when $m\ne 0$. 

It is convenient to paramaterize the strength of the flux line by the
positive quantity $\delta=e\Phi/(hc)$, where $e$ is the magnitude of
the charge on the particle and $\Phi$ is the total magnetic flux through
the singular point. In the presence of the flux line the Hamiltonian
is (see Brack {\it et al.} \cite{BBLMM95})
\begin{equation}
H=\frac{p_r^2}{2}+\frac{(p_\phi-\delta)^2}{2r^2}+V(r)
\label{flux1}
\end{equation}
where $p_\phi$, the momentum canonical to the coordinate $\phi$, is a
constant of the motion, and $V(r)$ is given by equation (\ref{eq27}).
Then, for the trajectory labelled by $j$ in equation (\ref{eq10}), the
angular part of the action is
\begin{equation}
S_{\rm ang}(\xi^{(j)};E) = p_\phi\xi^{(j)} = p_\phi(\gamma+2\pi j).
\label{flux2}
\end{equation}
Thus, the stationary phase condition (\ref{eq11}) becomes $p_\phi=m\hbar$.
However, by comparing (\ref{flux1}) with equation (\ref{eq3}) we see
that $(p_\phi-\delta)^2$ now plays the role that $L^2$ played in Sec.\ 
\ref{sec3}. For given $m$ the quantization condition (\ref{eq16a}) becomes
\begin{equation}
S_{\rm rad}(L=|m\hbar-\delta|;E)=2\pi\hbar(n_r+\sigma_m/4),\qquad\qquad
n_r=0,1,2,\dots
\label{flux3}
\end{equation}
Making this replacement in (\ref{eq29}) and setting $\sigma_m=2$
(corresponding to soft turnarounds at $r_-$ and $r_+$), we obtain the
following expression for the energy eigenvalues:
\begin{eqnarray}
&E_{mn_r}=\hbar\omega\{2n_r+[(m-\delta/\hbar)^2\pm\alpha^2/\hbar^2]^{1/2}+1\},
\nonumber\\
&m=0,\pm 1,\pm 2,\dots \qquad n_r=0,1,2,\dots
\label{flux4}
\end{eqnarray}

For the case of a singular flux line at the centre of a circular
harmonic oscillator potential, we can let $\alpha\to 0$ in (\ref{flux4}).
The resulting energy eigenvalues are
\begin{eqnarray}
&E_{mn_r}=\hbar\omega (2n_r+|m-\delta/\hbar|+1), \nonumber\\
&m=0,\pm 1,\pm 2,\dots \qquad n_r=0,1,2,\dots
\label{flux5}
\end{eqnarray}
which clearly are different for positive and negative values of $m$.
This result is the same as the exact analytical solution of the
Schr\"odinger equation for this problem, which can be obtained from
the exact harmonic oscillator solution by replacing $|m|\hbar$ by
$|m\hbar-\delta|$ wherever it occurs \cite{BBLMM95}.

\subsection{The circle billiard}
\label{sec4D}
In this section we apply the general formulation based on the transfer
operator to a particle moving in a constant potential (which we take
to be zero) inside a circle of radius $R$. We show that this leads to
well-known results for the EBK energy eigenvalues.

At energy $E$ and angular momentum $L$, the radial motion has an inner
turning point at the radius $r_-$ given by $|L|=r_-(2E)^{1/2}$. Choosing
the radius of the Poincar\'e circle to be $R$ (or just slightly less
than $R$ so that the PSS is crossed just after the particle has made a
collision with the boundary) ensures that all trajectories cross the PSS.

At fixed values of $E$ and $L$, the radial part of the action integral is
\begin{eqnarray}
&S_{\rm rad}(L;E)=2\int_{r_-}^R |p_r|dr=2\int_{r_-}^R(2E-L^2/r^2)^{1/2}dr
\nonumber\\
&=2(2ER^2-L^2)^{1/2} - 2|L|\cos^{-1}[|L|/(2E)^{1/2}R].
\label{eq31}
\end{eqnarray}
Writing $E=\hbar^2 k^2/2$ and setting $L=m\hbar$, with the values of $m$
restricted by the condition $|m|\hbar\le |L_{max}(E)|=(2E)^{1/2}R$, we
obtain from equation (\ref{eq16a}) the following condition for an
approximate energy eigenvalue of the quantum system:
\begin{eqnarray}
&(k^2 R^2-m^2)^{1/2}-|m|\cos^{-1}[|m|/(kR)]=\pi(n_r+{3\over 4}),
\nonumber\\
&m=0,\pm1,\pm2,\dots \qquad n_r=0,1,2,\dots
\label{eq32}
\end{eqnarray}
Here we have put $\sigma_m=3$, corresponding to a soft turnaround at
the inner turning point and a hard-wall collision at the circle boundary.
(See the discussion in Appendix \ref{appA}.) Equation (\ref{eq32}) can be
solved numerically to determine the EBK energy eigenvalues. (This is
equivalent to finding the zeros of the Bessel function $J_m(kR)$ when
it is approximated by the leading term of the Debye asymptotic expansion.
See Ref.\ \cite{BB97}, p 336.) Results
for the lowest energy eigenvalues have been tabulated by Keller and
Rubinow \cite{KR60} and by Brack and Bhaduri (see Ref.\ \cite{BB97}, p 88).
The fractional difference between the EBK eigenvalues and the exact
energy eigenvalues (determined from the zeros of the Bessel function
$J_m(kR)$) was found to decrease fairly rapidly with increasing energy.

\subsection{The annulus billiard}
\label{sec4E}
The annulus billiard consists of a particle moving in a constant
potential (which we take to be zero) in the region between two
concentric circles of radii $R$ and $a$. (We assume $R>a$.) In this
section we show how equation (\ref{eq16a}) may be used to obtain the
EBK energy eigenvalues for this system.

As in the case of the circle billiard, we choose the Poincar\'e circle
to have radius $R$. At energy $E$ and angular momentum $L$ the radial
motion may have an inner turning point at the radius $r_-$ given by
$|L|=r_- (2E)^{1/2}$. Provided $a<r_-$, the minimum value of $r$
during the radial motion will be $r_-$. However, if $a>r_-$, the
radial motion is reversed by a hard-wall collision at $r=a$. In the
former case with a soft turnaround at $r_-$, the Maslov index is
$\sigma_m=3$, as for the circle billiard. In the latter case, there
are two hard-wall collisions, and the Maslov index is $\sigma_m=4$.

The radial part of the action involves the same integral as in
equation (\ref{eq31}), but is now evaluated at the limits $r_{min}$
and $R$, where $r_{min}$ is $r_-$ or $a$, whichever is larger.
Putting $E=\hbar^2 k^2/2$ and setting $L=m\hbar$, with the values of
$m$ restricted by the condition $|m|\hbar\le|L_{max}(E)|=(2E)^{1/2}R$,
we obtain the following condition for an approximate energy eigenvalue
of the quantum system, similar to equation (\ref{eq32}):
\begin{eqnarray}
& \displaystyle (k^2 r^2-m^2)^{1/2}|_{r_{min}}^R
-|m|\cos^{-1}[|m|/(kr)]|_{r_{min}}^R =\pi(n_r+\sigma_m/4),
\nonumber\\
&m=0,\pm1,\pm2,\dots \qquad n_r=0,1,2,\dots
\label{eq33}
\end{eqnarray}
Here, $\sigma_m=3$ if $r_{min}=r_-$, and $\sigma_m=4$ if $r_{min}=a$.
The solution of this equation gives the EBK energy eigenvalue
$E_{mn_r}$. Numerical values for the lowest 30 distinct EBK energy
eigenvalues are tabulated in Ref.\ \cite{SG97} for three cases:
$a=0.1R$, $a=0.3R$, and $a=0.5R$. Also tabulated in Ref.\ \cite{SG97}
are the energy eigenvalues computed from Bogomolny's transfer operator
without making use of the stationary phase approximation or any other
approximation. The results differ appreciably from the EBK energy
eigenvalues. This draws attention to the fact that the stationary
phase approximation used to derive equation (\ref{eq16a}) constitutes
an approximation {\it in addition to} the main semiclassical
approximation contained in Bogomolny's transfer operator. It is clear
that Bogomolny's approach and EBK quantization are generally
equivalent only to leading order in $\hbar$.

\section {Three-dimensional systems with spherical symmetry}
\label{sec5}
A theory of the transfer operator for three-dimensional systems with
spherical symmetry is only slightly more complicated than in two
dimensions. For a particle of unit mass moving in a potential $V(r)$,
the Hamiltonian is 
\begin{equation}
H=\frac{p_r^2}{2} + \frac{p_\theta^2}{2r^2}
+\frac{p_\phi^2}{2r^2 \sin^2\theta} + V(r)
\label{eq4.1}
\end{equation}
where $p_r=\dot r$, $p_\theta=r^2\dot\theta$ and
$p_\phi=r^2\sin^2\theta\dot\phi$ are the momenta conjugate
to the spherical polar coordinates $(r,\theta,\phi)$ describing the
particle's position. Because the component of the angular momentum about
the polar axis, $p_\phi$, is a constant of the motion, let us denote it
as $L_z$. In addition, the square of the total angular momentum,
$L^2=p_\theta^2+L_z^2/\sin^2\theta$, is a constant of the motion.
Therefore, the Hamiltonian may be written as
\begin{equation}
H=\frac{p_r^2}{2} + \frac{L^2}{2r^2} + V(r)
\label{eq4.2}
\end{equation}
For given $E$ and $L$, the turning points of the radial motion are
determined by
\begin{equation}
\frac{L^2}{2r^2} + V(r) = E.
\label{eq4.3}
\end{equation}
Thus, as in two dimensions, the turning point radii, $r_-$ and $r_+$,
depend on $L^2$ as well as $E$.

In constructing the transfer operator, we choose the PSS to be a
sphere of radius $R$. The argument following equation (\ref{eq4})
applies equally well to the three-dimensional situation. Therefore, we
choose $R$ to be equal to the radius of the circular trajectory at
energy $E$ for which the angular momentum has its maximum possible value.
Then, for any angular momentum $|L|\le L_{max}(E)$, the trajectories of
the system will repeatedly cross this surface.

The generalized coordinates $q$ for the transfer operator are the
polar angles $(\theta,\phi)$ on the Poincar\'e sphere, which we denote
collectively as $\Omega$. Then, from equation (\ref{eq1}) with $f=3$,
the transfer operator from $\Omega$ to $\Omega^\prime$ is
\begin{equation}
T(\Omega^\prime,\Omega;E)=\sum_{\rm cl.tr.} {1\over{2\pi i\hbar}}
\left| {\rm det}{{\partial^2 S}\over
{\partial \Omega \partial \Omega^\prime}} \right|^{\frac{1}{2}}
\exp[iS(\Omega^\prime,\Omega;E)/\hbar-i\mu\pi/2],
\label{eq4.4}
\end{equation}
Here the sum is over all possible classical trajectories at energy
$E$ which go from $\Omega$ to $\Omega^\prime$ without crossing the
PSS (in the same sense) at any other point. Let $\gamma$ be the
angle subtended at the origin by these points, with $0\le\gamma\le\pi$.
All possible trajectories from $\Omega$ to $\Omega^\prime$ lie in the
plane defined by these points and the origin, and each such trajectory
can be uniquely identified by the total angle traversed as the
particle moves along the trajectory. The trajectories fall into two
classes, which we treat separately. In the first class, the
angles traversed are $\xi_+^{(j)}=2\pi j+\gamma$, $j=0,1,2,\dots$,
while in the second class, the angles traversed are $\xi_-^{(j)}
=2\pi j-\gamma$, $j=1,2,\dots$. The sum in (\ref{eq4.4}) is over the
trajectories of both classes.

The determinant of the second derivatives of $S(\xi_{\pm}^{(j)};E)$ is
evaluated in Appendix \ref{appB}, where it is found that
\begin{equation}
\left| {\rm det}{{\partial^2 S}\over{\partial\Omega \partial\Omega^\prime}}
\right| = {{1}\over{\sin\gamma}} \left| {{\partial S}\over{\partial\gamma}}
\right|\, \left| {{\partial^2S}\over{\partial\gamma^2}} \right|.
\label{eq4.5}
\end{equation}
Substituting in equation (\ref{eq4.4}), we obtain
\begin{equation}
T(\Omega^\prime,\Omega;E)=\sum_{cl.tr.} {1\over{2\pi i\hbar}}
{{1}\over{(\sin\gamma)^{1/2}}}
\left| {{\partial S}\over{\partial\gamma}} \right|^{\frac{1}{2}}
\, \left| {{\partial^2S}\over{\partial\gamma^2}} \right|^{\frac{1}{2}}
\exp[iS(\xi_{\pm}^{(j)};E)/\hbar-i\mu_{\pm}^{(j)}\pi/2],
\label{eq4.6}
\end{equation}

The spherical symmetry of the system implies that
$T(\Omega^\prime,\Omega;E)$ depends only on the angle $\gamma$ subtended
at the origin by the points $\Omega$ and $\Omega^\prime$ on the Poincar\'e
sphere. Therefore, we can expand the transfer operator in a Legendre series:
\begin{equation}
T(\Omega^\prime,\Omega;E) = T(\gamma;E)
=\sum_{l=0}^\infty C_l(E)\,P_l(\cos\gamma),
\label{eq4.7}
\end{equation}
where the expansion coefficients are
\begin{equation}
C_l(E)=\displaystyle\frac{2l+1}{2}\int_0^\pi T(\gamma;E)\,P_l(\cos\gamma)
\sin\gamma \, d\gamma.
\label{eq4.8}
\end{equation}
The spherical harmonic addition theorem
\begin{equation}
P_l(\cos\gamma)=\displaystyle \frac{4\pi}{2l+1} \sum_{m=-l}^l
Y_{lm}(\Omega) Y_{lm}^\ast(\Omega^\prime),
\label{eq4.9}
\end{equation}
allows us to write
\begin{equation}
T(\Omega^\prime,\Omega;E) = T(\gamma;E) = \displaystyle
\sum_{l=0}^\infty C_l(E) \frac{4\pi}{2l+1}
\sum_{m=-l}^l Y_{lm}(\Omega) Y_{lm}^\ast(\Omega^\prime),
\label{eq4.10}
\end{equation}
where $Y_{lm}(\Omega)$ is a spherical harmonic function. We can now
calculate matrix elements of the $T$-operator in the angular momentum
representation. From the orthonormality of the spherical harmonics,
a general matrix element is found to be
\begin{eqnarray}
T_{l_1m_1,l_2m_2}(E)&=\int d\Omega \int d\Omega^\prime
Y_{l_1m_1}^\ast(\Omega)\,T(\Omega^\prime,\Omega;E)\,Y_{l_2m_2}(\Omega^\prime)
\nonumber\\
&=\displaystyle \left(\frac{4\pi}{2l_1+1}\right) C_{l_1}(E)
\delta_{l_1l_2}\delta_{m_1m_2}.
\label{eq4.11}
\end{eqnarray}
Thus, the $T$-matrix is diagonal in this representation, and its
eigenvalues (as a function of $E$) are just the diagonal elements. Denoting
the $lm$-eigenvalue curve, which is $(2l+1)$-fold degenerate, as
$\lambda_{lm}(E)$, we obtain from equations (\ref{eq4.6}), (\ref{eq4.8})
and (\ref{eq4.11}),
\begin{equation}
\lambda_{lm}(E) = \displaystyle \frac{1}{i\hbar} \sum_{cl.tr.} \int_0^\pi
d\gamma \left| \frac{\partial S}{\partial\gamma} \right|^{\frac{1}{2}}
\left| \frac{\partial^2 S}{\partial\gamma^2} \right|^{\frac{1}{2}}
\exp[iS(\xi_{\pm}^{(j)};E)/\hbar-i\mu_{\pm}^{(j)}\pi/2]
\,P_l(\cos\gamma)(\sin\gamma)^{1/2}.
\label{eq4.12}
\end{equation}

So far we have made no approximations beyond those used to derive
Bogomolny's semiclassical transfer operator. For a given system, one
can calculate the action $S(\xi_{\pm}^{(j)};E)$ for each possible
trajectory, as well as the derivatives with respect to $\gamma$,
and hence evaluate the integrals. However, to proceed in a manner
analogous to our treatment of two-dimensional systems, we make use of
the following asymptotic expansion for $P_l(\cos\gamma)$, valid for
large values of $l$ (see Ref.\ \cite{AS65}, equations 8.10.7 and 6.1.46):
\begin{equation}
P_l(\cos\gamma) \approx \displaystyle {{1}\over{(l+\frac{1}{2})^{1/2}}}
\left( \frac{1}{2\pi\sin\gamma} \right)^{1/2}
\{ \exp[i(l+\mbox{${1\over 2}$})\gamma-i\pi/4]
 + \exp[-i(l+\mbox{${1\over 2}$})\gamma+i\pi/4] \},
\label{eq4.13}
\end{equation}
where terms of higher order in $1/l$ have been neglected. The same
approximation has been used in the problem of scattering of waves by a
sphere to show how the limit of geometrical optics can be obtained
from physical optics \cite{Newton82}. After inserting (\ref{eq4.13})
in equation (\ref{eq4.12}), we can evaluate the resulting integrals
using the stationary phase approximation.

The two terms in (\ref{eq4.13}) yield the stationary phase conditions
\begin{equation}
\displaystyle \frac{\partial S(\xi_{\pm}^{(j)};E)}{\partial\gamma}
\pm(l+\mbox{${1\over 2}$})\hbar=0,
\label{eq4.14}
\end{equation}
where the $\pm$ between the terms refers to the two terms in the
asymptotic expansion
(\ref{eq4.13}). At this point we find it convenient to treat separately
the trajectories that traverse angles $\xi_+^{(j)}=2\pi j+\gamma$,
$j=0,1,2,\dots$ and those that traverse angles $\xi_-^{(j)}=2\pi j-\gamma$,
$j=1,2,\dots$, where in both cases $0\le\gamma\le \pi$.
We shall show that both classes of trajectory lead to the
same stationary phase condition, but that for a given value of $l$,
the corresponding trajectory is either in one class or the other.

Let us suppose for the moment that the trajectory under consideration
belongs to the first class. Because the action increases
with the traversed angle $\xi_+^{(j)}$, we have
$\partial S/\partial\gamma$$=$$\partial S/\partial\xi_+^{(j)} >0$.
Then equation (\ref{eq4.14}) can be satisfied by taking the second
term in (\ref{eq4.13}). Since $\partial S/\partial\gamma$ is the
magnitude of the total angular momentum, the stationary phase condition
(\ref{eq4.14}) effectively quantizes the total angular momentum:
\begin{equation}
L=|{\bf L}|=(l+\mbox{${1\over 2}$})\hbar,\qquad\qquad l=0,1,2,\dots
\label{eq4.15}
\end{equation}
Furthermore, specifying the total angular momentum (with $L\le L_{max}(E)$;
see below equation (\ref{eq4.3})) completely determines the trajectory at
energy $E$ in the plane containing $\Omega$, $\Omega^\prime$ and the origin.
Thus, for a given value of $l$ in (\ref{eq4.15}), there is {\it at most
one trajectory}, labelled by $j_l\ge 0$, which contributes to the sum
in equation (\ref{eq4.12}). For this trajectory (which, by assumption,
belongs to the first class), we denote the angle traversed by the particle
as $\xi_l=\gamma_l+2\pi j_l$, where $\gamma_l$ (in the range between 0
and $\pi$) is the stationary point determined by equation (\ref{eq4.14}).
We also denote the corresponding action by $S_l$ and the phase index
by $\mu_l$. When slowly varying quantities are taken outside the
integral (to be evaluated at $\gamma=\gamma_l$), the $T$-matrix
eigenvalue curve (\ref{eq4.12}) becomes
\begin{eqnarray}
&\lambda_{lm}(E)=\displaystyle\frac{1}{(2\pi)^{1/2} i\hbar}
\frac{1}{(l+{1\over 2})^{1/2}}\left|\frac{\partial S}{\partial\gamma}
\right|^{\frac{1}{2}}_{\gamma=\gamma_l} \left| \frac{\partial^2 S}
{\partial\gamma^2} \right|^{\frac{1}{2}}_{\gamma=\gamma_l}\nonumber\\
&\displaystyle \times \int_0^{\pi} d\gamma\,\exp[iS_l(\gamma;E)/\hbar
-i(l+\mbox{${1\over 2}$})\gamma-i\mu_l\pi/2 +i\pi/4]
\label{eq4.16}
\end{eqnarray}

The integral is now evaluated in the usual way.
Introducing the phase index $\nu_l$ as in equation (\ref{eq12}), we obtain
\begin{equation}
\lambda_{lm}(E) \approx \exp[iS_l(\xi_l;E)/\hbar
-i(l+\mbox{${1\over 2}$})\gamma_l-i(\mu_l+\nu_l)\pi/2]
\label{eq4.17}
\end{equation}
As in the two-dimensional case, these approximate $T$-matrix eigenvalue
curves have unit modulus, consistent with the $T$-matrix being unitary.

We now split the action $S(\xi_l;E)$ into radial and
angular parts. From the stationary phase condition, equation
(\ref{eq4.14}), the magnitude of the angular momentum along the
trajectory is $(l+\frac{1}{2})\hbar$. Thus, since the particle traverses
the angle $\xi_l=\gamma_l+2\pi j_l$ when it moves along this trajectory,
we obtain for the angular part of the action
\begin{equation}
S_{\rm ang}(\xi_l;E)/\hbar=(l+\mbox{${1\over 2}$})\,(\gamma_l+2\pi j_l).
\label{eq4.18}
\end{equation}
The radial part of the action, evaluated for $L^2=(l+\frac{1}{2})^2
\hbar^2$, is denoted as
\begin{equation}
S_{\rm rad}(L^2=(l+\mbox{${1\over 2}$})^2\hbar^2;E)
=\oint |p_r| |dr|=2\int_{r_-}^{r_+}|p_r|dr. 
\label{eq4.19}
\end{equation}
Hence, equation (\ref{eq4.17}) becomes
\begin{equation}
\lambda_{lm}(E)\approx\exp[iS_{\rm rad}(L^2=(l+\mbox{${1\over 2}$})^2
\hbar^2;E)/\hbar-i(\mu_l+\nu_l-2j_l)\pi/2].
\label{eq4.20}
\end{equation}

Next we suppose that the trajectory corresponding to a given value
of $l$ belongs to the second class, in which case the angle traversed
by the particle has the form $\xi_-^{(j)}=2\pi j-\gamma$, $j=1,2,\dots$
Here too the action must increase with the traversed angle $\xi_-^{(j)}$,
implying that $\partial S/\partial\gamma$ $=$ $-\partial
S/\partial\xi_-^{(j)}<0$. Then the stationary phase condition, equation
(\ref{eq4.14}), can be satisfied by taking the first term in the
asymptotic expansion (\ref{eq4.13}). One obtains in this case the same
condition, equation (\ref{eq4.15}), for the magnitude of the angular
momentum. However, the $T$-matrix eigenvalue curve corresponding to
equation (\ref{eq4.17}) becomes in this case
\begin{equation}
\lambda_{lm}(E) \approx \exp[iS_l(\xi_l;E)/\hbar+i(l+\mbox{${1\over 2}$})
\gamma_l - i(\mu_l+\nu_l+1)\pi/2],
\label{eq4.21}
\end{equation}
where $\xi_l=2\pi j_l-\gamma_l$ and $j_l\ge1$. The angular part of the
action in this case is
\begin{equation}
S_{\rm ang}(\xi_l;E)/\hbar=(l+\mbox{${1\over 2}$})\,(2\pi j_l-\gamma_l),
\label{eq4.22}
\end{equation}
which yields the $T$-matrix eigenvalue curve
\begin{equation}
\lambda_{lm}(E)\approx\exp[iS_{\rm rad}(L^2=(l+\mbox{${1\over 2}$})^2
\hbar^2;E)/\hbar - i(\mu_l+\nu_l-2j_l+1)\pi/2].
\label{eq4.23}
\end{equation}

The semiclassical energy eigenvalues of the quantum system are found
from the determinantal equation (\ref{eq2}), which is satisfied
whenever an eigenvalue of the $T$-matrix is equal to unity. Thus, the
condition for an energy eigenvalue is that $\lambda_{lm}(E)=\exp(i2\pi n_r)$.
>From equations (\ref{eq4.20}) and (\ref{eq4.23}) this yields the following
condition for an energy eigenvalue:
\begin{equation}
S_{\rm rad}(L^2=(l+\mbox{$\frac{1}{2}$})^2\hbar^2;E)
=2\pi\hbar(n_r+\sigma_l/4), \qquad\qquad n_r=0,1,2\dots
\label{eq4.24}
\end{equation}
where the Maslov index $\sigma_l$ associated with a complete cycle of
the radial motion at energy $E$ and $L^2=(l+1/2)^2\hbar^2$ is defined
to be
\begin{eqnarray}
&\sigma_l=\mu_l+\nu_l-2j_l \qquad\qquad {\rm for\ trajectories\ in\ class}
\qquad \xi_l=2\pi j_l+\gamma_l, \qquad j_l=0,1,\dots \nonumber\\
&\sigma_l=\mu_l+\nu_l-2j_l+1 \qquad {\rm for\ trajectories\ in\ class}
\qquad \xi_l=2\pi j_l-\gamma_l, \qquad j_l=1,2,\dots
\label{eq4.25}
\end{eqnarray}
As in the two-dimensional case, the values of $n_r$ in (\ref{eq4.24})
are determined by the assumption that $S_{\rm rad}\ge 0$.

It is shown in Appendix \ref{appA} that this definition always leads to
the result $\sigma_l=2$ for soft potentials, and to $\sigma_l=3$ for a
particle inside a spherical cavity, making a hard-wall collision
with the boundary in each cycle of the radial motion. These results are
consistent with computing $\sigma_l$ by the simple EBK rules of counting
1 for each soft turnaround of the radial motion and 2 for each hard-wall
collision. Thus, equations (\ref{eq4.24}) and (\ref{eq4.25}) give the EBK
quantization condition for the radial part of the action in three
dimensions. The other EBK quantization conditions are
$L^2=(l+\frac{1}{2})^2\hbar^2$ and $L_z=m\hbar$ with $-l\le m\le l$.
The first of these was obtained earlier from the stationary phase condition.
The second is a natural interpretation of the quantum number $m$ introduced
through the expansion of the $T$-operator in spherical
harmonics. Because of the spherical symmetry, the energy eigenvalues
corresponding to a particular value of $l$ are $(2l+1)$-fold degenerate.

\section{Application to systems with spherical symmetry}
\label{sec6}
\subsection{The Coulomb plus $1/r^2$ potential in three dimensions}
\label{sec6A}
As in the two-dimensional case, the pure Coulomb potential has the
special property that all classical trajectories (ellipses) starting
from a given point on the PSS will return to the same point, which is,
therefore, a focal point. To avoid this singular behaviour we shall
again add a small $1/r^2$ term to the potential, which may be
attractive or repulsive. Thus, we take the Hamiltonian corresponding
to equation (\ref{eq4.2}) to be
\begin{equation}
H=\frac{p_r^2}{2}-\frac{1}{r}+ \frac{L^2\pm\alpha^2}{2r^2},
\label{eq4B.1}
\end{equation}
where, as in the two-dimensional case, $\alpha^2/2$ is the strength of
the $1/r^2$ potential. The turning-point radii, determined from equation
(\ref{eq4.3}), are given by equations (\ref{eq18}) and (\ref{eq19}) with
$L^2$ being the square of the angular momentum of the particle.
As in the two-dimensional case, we choose the radius of the Poincar\'e
sphere to be $R=1/(2|E|)$.

The calculation of the radial contribution to the action proceeds
exactly as in two dimensions. The result, similar to equation (\ref{eq21}),
is
\begin{equation}
S_{\rm rad}(L^2=(l+\mbox{${1\over 2}$})^2 \hbar^2;E)
=\displaystyle \pi \left( \frac{2}{|E|} \right)^{1/2}
-2\pi [(l+\mbox{${1\over 2}$})^2\hbar^2\pm\alpha^2]^{1/2}.
\label{eq4B.2}
\end{equation}
Setting the Maslov index $\sigma_l$ equal to 2, corresponding to two
soft turnarounds in the radial motion (see Appendix A), we obtain from
equation (\ref{eq4.24}) the following condition for an energy eigenvalue
for given $l$ and $m$:
\begin{equation}
\displaystyle \pi \left( \frac{2}{|E|} \right)^{1/2}
-2\pi [(l+\mbox{${1\over 2}$})^2\hbar^2\pm\alpha^2]^{1/2}
=2\pi\hbar(n_r+\mbox{${1\over 2}$}), \qquad\qquad n_r=0,1,2,\dots
\label{eq4B.3}
\end{equation}
Since $E$ is negative for the bound-state solutions we are considering,
the energy eigenvalues for given $l$ and $m$ ($-l\le m\le l$) are
found to be
\begin{equation}
E_{lmn_r} =-{{1}\over{2\hbar^2 \{ n_r+{1\over 2} +
[(l+{1\over 2})^2\pm\alpha^2/\hbar^2]^{1/2} \}^2}},\qquad\qquad 
n_r=0,1,2,\dots
\label{eq4B.4}
\end{equation}
This expression gives the approximate semiclassical energy eigenvalues
for the Coulomb plus $1/r^2$ potential. As expected, they are
$(2l+1)$-fold degenerate because of the spherical symmetry of the system.
It should also be observed that the allowed
values of $l$ are constrained by the condition that
$(l+\frac{1}{2})\hbar\le L_{max}(E)$, with
$[L_{max}(E)]^2=1/(2|E|)\mp\alpha^2$,
as in the two-dimensional case.

The result for the pure Coulomb potential is found by letting
$\alpha\to 0$ in equation (\ref{eq4B.4}). Putting $n=l+n_r+1$ we
obtain,
\begin{equation}
E_n= -{{1}\over{2\hbar^2 n^2}},\qquad\qquad n=1,2,\dots
\label{eq4B.5}
\end{equation}
independent of the sign of the $\alpha^2$ term, as in two dimensions.
This is the same as the familiar result found by solving the
three-dimensional Schr\"odinger equation for the Coulomb potential.
It is clear from the definition of $n$ that $l<n$. This condition,
which also arises in solving the Schr\"odinger equation, leads to the
degeneracy of the $n$th energy level being $n^2$.

It is remarkable that we have obtained the same result as the exact
energy levels of the hydrogen atom, despite having made three
approximations: (i) the semiclassical approximation embodied in
Bogomolny's transfer operator; (ii) the asymptotic expansion (\ref{eq4.13})
for $P_l(\cos\gamma)$; and (iii) the evaluation of the integral in
(\ref{eq4.12}) using the stationary phase approximation. A plausible
explanation for this agreement will be given in the discussion at the
end of the paper.

\subsection{The spherical harmonic oscillator plus $1/r^2$ potential}
\label{sec6B}
In this subsection we treat the isotropic harmonic oscillator plus a
small $1/r^2$ potential.  We take
the Hamiltonian corresponding to equation (\ref{eq4.2}) to be
\begin{equation}
H=\frac{p_r^2}{2}+\frac{1}{2}\omega^2 r^2 + \frac{L^2\pm\alpha^2}{2r^2},
\label{eq4C.1}
\end{equation}
where, as in the two-dimensional case, $\omega^2$ describes the
steepness of the harmonic oscillator potential, and $\alpha^2/2$ is
the strength of the $1/r^2$ potential, which may be attractive or repulsive.
The radial turning points, determined from equation (\ref{eq4.3}) for
fixed $E$ and $L^2$, are given by equation (\ref{eq28}), as in two dimensions.
When $L^2$ has its maximum possible value, determined by
\begin{equation}
[L_{max}(E)]^2={{E^2}\over{\omega^2}}\mp\alpha^2,
\label{eq4C.3}
\end{equation}
the radial kinetic energy is zero and the particle trajectory is confined
to the sphere of radius $R=E^{1/2}/\omega$. We take the Poincar\'e sphere
to have this radius since all trajectories having $L\le L_{max}(E)$
must repeatedly cross this surface.

For given $E$ and $L^2$ the radial part of the action can be calculated
as in the two-dimensional case. The result is the same as equation
(\ref{eq29}). Furthermore, from the analysis in Appendix A, the Maslov index
is $\sigma_l=2$. Thus, from equation (\ref{eq4.24}), the condition for
an energy eigenvalue is
\begin{equation}
{{\pi E}\over{\omega}}-\pi\hbar[(l+\mbox{${1\over 2}$})^2
\pm\alpha^2/\hbar^2]^{1/2}
=2\pi\hbar(n_r+\mbox{${1\over 2}$}),\qquad\qquad n_r=0,1,2\dots
\label{eq4C.5}
\end{equation}
Hence, the energy eigenvalue specified by $l$, $m$ and $n_r$ is
\begin{equation}
E_{lmn_r}= \hbar\omega \{ 2n_r+[(l+\mbox{${1\over 2}$})^2
\pm\alpha^2/\hbar^2]^{1/2} +1 \},\qquad\qquad n_r=0,1,2,\dots
\label{eq4C.6}
\end{equation}
These eigenvalues are clearly $(2l+1)$-fold degenerate. Note that the
permissible values of $l$ are determined by the condition that
$(l+\frac{1}{2})\hbar\le L_{max}(E)$, with $L_{max}(E)$ given by equation
(\ref{eq4C.3}).

The energies of the pure isotropic harmonic oscillator in three
dimensions are found by letting $\alpha\to 0$. Defining $n=2n_r+l$, we
obtain (independent of the sign of the $\alpha^2$ term)
\begin{equation}
E_n=\hbar\omega(n+\mbox{${3\over 2}$}), \qquad\qquad n=0,1,2,\dots
\label{eq4C.7}
\end{equation}
The ground state energy is the zero-point energy associated with 3
freedoms. The multiplicities of the levels are determined by the
$(2l+1)$-fold degeneracy associated with $m$ and by the number of distinct
ways of obtaining a given value of $n$ from the integer values of $l$
and $n_r$. The energy levels given by (\ref{eq4C.7}) and their
degeneracies are the same as those obtained in the solution of the
three-dimensional Schr\"odinger equation for the spherical oscillator
(see, for example, Ref.\ \cite{Flugge74}, pp 166-168).

\subsection{Billiard in a spherical cavity}
\label{sec6C}
A particle moving in zero potential inside a spherical cavity of
radius $R$ is the three-dimensional analogue of the circle billiard. As
in the two-dimensional system, for given values of the energy $E$ and
the square of the total angular momentum $L^2$, the inner turning
point radius $r_-$ is given by $|L|=(2E)^{1/2}r_-$. We choose the PSS
to be the sphere of radius $R$ (or just slightly less than $R$, in order
that the trajectories cross the PSS immediately after colliding with
the spherical boundary).

To obtain the EBK eigenvalues from equation (\ref{eq4.24})
we can use equation (\ref{eq31}) for the radial part of
the action integral, which is the same for the two and
three-dimensional systems. Furthermore, it was shown
in Appendix A that the Maslov index is $\sigma_l=3$ for this system.
Thus, from equations (\ref{eq4.24}) and(\ref{eq31}), the condition for
an energy eigenvalue of the billiard in a spherical cavity is
\begin{eqnarray}
&[k^2 R^2-(l+{1\over 2})^2]^{1/2}-(l+{1\over 2})\cos^{-1}[(l+{1\over 2})/(kR)]
=\pi(n_r+{3\over 4}), \nonumber\\
& \qquad l=0,1,2\dots, \qquad n_r=0,1,2,\dots
\label{eq4D.1}
\end{eqnarray}
Here the allowed values of $l$ at energy $E$ are restricted by the
condition $(l+\frac{1}{2})\le kR$, where $k=(2E)^{1/2}/\hbar$. Equation
(\ref{eq4D.1}) could be solved numerically in a manner similar to
equation (\ref{eq32}) to obtain the EBK energy eigenvalues. (This would
be equivalent to finding approximate values for the zeros of the
spherical Bessel function $j_m(kR)$ when it is approximated by the
leading term of the Debye asymptotic expansion.)

\section{Discussion}
\label{discussion}
We have shown how, with the help of the stationary phase
approximation, Bogomolny's transfer operator leads to the EBK
quantization rules for the energy eigenvalues of integrable systems
having rotational symmetry, in both two and three dimensions.
An important aspect of the theory was showing that the Maslov indices
are correctly given by the simple rules of counting 1 for each soft
turnaround and 2 for each hard-wall collision occurring during one
complete cycle of the radial motion.

In discussing the annulus billiard in Sec.\ \ref{sec4E} we drew
attention to the fact that the EBK energy eigenvalues are appreciably
different from those obtained using Bogomolny's transfer operator
unmodified by making the stationary phase approximation or any other
approximation. It is noteworthy, however, that among the eight different
rotationally invariant systems to which the theory was applied, in five
cases the EBK energy eigenvalues turned out to be the same as those obtained
from an exact solution of the Schr\"odinger equation. In fact, for the
hydrogen atom in three dimensions (Sec.\ \ref{sec6A}), it was pointed out
that the correct energy levels were obtained despite having made three
significant approximations in the theoretical development. How can one
understand this surprising result?

A similar situation arose in a recent paper \cite{TG97} concerning the
application of Bogomolny's $T$-operator to a circular harmonic
oscillator plus $1/r^2$ potential. For that system it was possible to
write down an exact transfer operator, which led to the exact energy
eigenvalues without making any approximations.  It was then shown,
with the help of the Poisson summation formula, that the exact
transfer operator could be written as an infinite sum of certain
integrals, and that the leading term in this sum was the same as
Bogomolny's transfer operator. Further, it was shown that improving
the stationary phase approximation leads to corrections to the energy
eigenvalues that involve higher powers of $\hbar$ than the leading
term. In short, the semiclassical result agreed with the exact quantum
result simply because corrections to the semiclassical approximation
(which are nonzero) and to the stationary phase approximation (which
are also nonzero) were not evaluated. Presumably, if corrections to
all the approximations were evaluated systematically, they would cancel
each other in all orders of $\hbar$.

It seems likely that we are dealing with a similar situation for the
hydrogen atom in two and three dimensions, and for the harmonic
oscillator with a singular magnetic flux line. For these systems it does
not appear to be easy to write down an exact transfer operator, so
that we are unable to carry out an analysis similar to that of Ref.\
\cite{TG97}. Nevertheless, if the EBK energy eigenvalue is regarded as the
leading term of a semiclassical expansion in increasing powers of $\hbar$,
it is perhaps not so surprising that, in some cases, it agrees with the
exact quantum result.

\acknowledgments
The work reported in this paper was stimulated by research carried out
earlier by Julie Lefebvre. We are greatly indebted to her and to
Rajat Bhaduri, Stephen Creagh, Donald Sprung and 
Peiqing Tong for helpful discussions. Support from the Natural Sciences
and Engineering Research Council of Canada is gratefully acknowledged.

\appendix

\section{Equivalence of $\sigma$ with the Maslov index
in EBK quantization}
\label{appA}
In this Appendix we consider the index $\sigma_m=\mu_m+\nu_m$ introduced
following equation (\ref{eq16a}) for a particle in two dimensions,
and the index $\sigma_l=\mu_l+\nu_l$ defined in equation (\ref{eq4.25})
for a particle in three dimensions. In both cases we consider a particle
confined inside a finite region either by a smooth potential or a hard-wall
boundary. From the definitions of the $\mu$ and $\nu$ indices we show that
$\sigma_m$ and $\sigma_l$ have the values one would compute from the
simple rules for obtaining the Maslov index in EBK quantization.

For two-dimensional systems, consider a particle moving in a smooth
potential $V(r)$. Figure 1 shows typical trajectories along with
trajectories having slightly larger and slightly smaller angular
momenta than the main trajectory.
The trajectory with smaller angular momentum approaches more
closely to the origin, but on the outward leg also ventures farther away
from the origin. In the course of doing so, it must necessarily
intersect the trajectory with larger angular momentum. Thus, there is
a focal point and $\mu_m$ is thereby incremented by 1. There are then two
possibilities. Before returning to the Poincar\'e circle, the two orbits
may intersect yet again, leading to $\mu_m=2$, as shown in Fig.\ 1(a).
If this happens one observes that the angle traversed is greater for
the larger angular momentum, so that $\partial^2 S /\partial\gamma^2> 0$
and $\nu_m=0$. On the other hand, there may be no further intersection so
that $\mu_m=1$, as illustrated in Fig.\ 1(b). In this case one observes
that $\partial^2 S /\partial\gamma^2< 0$ so that $\nu_m=1$. In either
event we have $\sigma_m=2$, which is the same as the value for the Maslov
index in EBK quantization obtained by counting 1 for each of the
turning points of the radial motion.

We have performed numerical studies for two-dimensional systems having
potentials of the form $V(r)\sim\pm r^k$, where the plus sign is
assumed for $k>0$ and the minus sign for $k<0$ (so that the potential
is attractive). Our calculations
show that for $-1<k< 2$ one finds $\mu_m=2$ and $\nu_m=0$, while
for $-2<k<-1$ and $2<k<\infty$ one finds $\mu_m=1$ and $\nu_m=1$.
The cases $k=-1$ and $k=2$, which are self-focusing, are marginal for
the present analysis.

The case of a hard wall may be approximated by taking the limit
$k\rightarrow\infty$. However,
when considering both large $k$ and infinitesimally close trajectories,
one must be careful about the order in which the limits are taken.
For any fixed $k$, no matter how large, we can take the two nearby
angular momenta close enough to the main trajectory that we obtain
$\sigma_m=2$, in conformity with the previous discussion. On the other
hand, if we consider two fixed angular momenta, no matter how close,
we can make $k$ large enough that the two trajectories no longer
intersect on the outer loop. This is the appropriate analysis for a
disk with infinitely hard walls. In this case $\mu_m=2$ from the
collision with the hard wall (Dirichlet boundary conditions), while
$\nu_m=1$, giving the result $\sigma_m=3$. This agrees with the result
for the EBK Maslov index obtained by counting 1 for the soft turnaround
at the inner turning point and 2 for the hard turnaround at the disk
boundary. The same analysis applies even if the motion is not
force-free within the disk. For example, one could include a uniform
magnetic field, or a flux line, or a harmonic potential out to the
disk radius. As long as the trajectory bounces off the disk boundary,
$\sigma_m=3$, but if the particle avoids colliding with the boundary,
the discussion in the previous paragraph will apply and $\sigma_m=2$.

Turning now to three-dimensional systems with a smooth potential,
we recall that the Maslov index $\sigma_l$ was defined in equation
(\ref{eq4.25}) for the two different classes of trajectories. We now
show that, for soft potentials, this leads to the result $\sigma_l=2$
in all cases.

For given points $\Omega$ and $\Omega^\prime$ on the
Poincar\'e sphere, a trajectory from $\Omega$ to $\Omega^\prime$ lies
in the plane containing these points and the origin. Focal points
arising from variations within the plane of this trajectory yield the
result $\mu_l+\nu_l=2$ from the preceding analysis. In three dimensions,
however, there may exist an additional focal point lying on the
straight line from the point $\Omega$ to the origin. One can see that,
provided the particle traverses an angle between $\pi$ and $2\pi$ in
moving along the trajectory for specified $E$ and $L^2$, it will
intersect this line at a point on the opposite side of the origin from
$\Omega$. Similar trajectories in all planes containing $\Omega$ and
the origin will intersect at this point, which is, therefore, a focal
point. In this case, $\mu_l$ must be incremented by 1, giving
$\mu_l+\nu_l=3$.  But the angle traversed by the particle in this
case is $\xi_l=2\pi-\gamma_l$, which means the trajectory belongs to the
class $\xi_l=2\pi j_l-\gamma_l$ with $j_l=1$. Hence, from equation
(\ref{eq4.25}), the Maslov index is $\sigma_l=2$. Note that if the
particle makes an additional circuit around the origin (corresponding
to $\xi_l=2\pi j_l-\gamma_l$ with $j_l=2$) there will be two additional 
focal points on the line joining $\Omega$ to the origin (on both sides
of the origin), but from (\ref{eq4.25}) the Maslov index will still be
$\sigma_l=2$. Clearly this generalizes to any number of complete circuits
around the origin within the time of one cycle of the radial motion.

If the angle traversed by the particle in moving along the
trajectory specified by $E$ and $L^2$ lies between 0 and $\pi$, there
is no focal point of the type described in the preceding paragraph.
The angle traversed is $\xi_l=\gamma_l+2\pi j_l$ with $j_l=0$, and
equation (\ref{eq4.25}) gives $\sigma_l=2$. If there are additional
circuits around the origin corresponding to $\xi_l=\gamma_l+2\pi j_l$,
with $j_l=1,2,\dots$, one can easily see that $\mu_l$ must be
incremented by 2 for each circuit, but from equation (\ref{eq4.25}),
the Maslov index remains $\sigma_l=2$. Thus, we have shown that
$\sigma_l=2$ for all possible trajectories in a soft potential. This
result is the same as would be obtained from the simple EBK rule of
counting 1 for each radial turning point of the effective potential
$V(r)+L^2/(2r^2)$.

If the outer radial turning point is replaced by a hard-wall collision
with the boundary (Dirichlet boundary condition on the wave function),
the above analysis still holds up to the point of colliding with the
boundary. As for the two-dimensional systems, the hard-wall collision
requires incrementing $\mu_l$ by 1, giving the result $\sigma_l=3$.
This is the EBK result from the simple rule of counting 1 for the soft
turnaround at the inner radial turning point plus 2 for the collision
with the boundary.

\section{The determinant of second derivatives of $S$}
\label{appB}
In this appendix we evaluate the determinant of second derivatives of
the action, which constitutes the amplitude of the three-dimensional
transfer operator:
\begin{equation}
{\rm det}{{\partial^2 S}\over{\partial\Omega\partial\Omega^\prime}}
=\left| \matrix{
\displaystyle {{\partial^2S}\over{\partial\theta\,\partial\theta^\prime}} &
\displaystyle {{\partial^2S}\over{\partial\theta\,\partial\phi^\prime}} \cr
\displaystyle {{\partial^2S}\over{\partial\phi\,\partial\theta^\prime}} &
\displaystyle {{\partial^2S}\over{\partial\phi\,\partial\phi^\prime}} \cr
} \right|.
\label{B1}
\end{equation}
Here $S=S(\xi_{\pm}^{(j)};E)$, which, through $\xi_{\pm}^{(j)}$,
depends on the angle $\gamma$ introduced just after equation (\ref{eq4.4}).
Since $\gamma$ is the angle subtended at the origin by the
points $\Omega$ and $\Omega^\prime$ on the Poincar\'e sphere, we have
\begin{equation}
\cos\gamma = \sin\theta\,\sin\theta^\prime\,\cos(\phi-\phi^\prime)
+\cos\theta\,\cos\theta^\prime,\qquad\qquad 0\le\gamma\le\pi.
\label{B2}
\end{equation}
Our objective is to express the determinant in terms of $\gamma$.

First, let us write
\begin{eqnarray}
&\displaystyle {{\partial S}\over{\partial\theta}}=
{{\partial S}\over{\partial\gamma}}\,\,
{{\partial \gamma}\over{\partial\theta}} \nonumber\\
&\displaystyle{{\partial^2S}\over{\partial\theta\,\partial\theta^\prime}}=
{{\partial^2S}\over{\partial\gamma^2}}\,\,
{{\partial\gamma}\over{\partial\theta}}\,\,
{{\partial\gamma}\over{\partial\theta^\prime}}
+{{\partial S}\over{\partial\gamma}}\,\,
{{\partial^2\gamma}\over{\partial\theta\,\partial\theta^\prime}}.
\label{B3}
\end{eqnarray}
It will greatly simplify the calculation to evaluate the second
derivatives assuming that $\theta=\theta^\prime=\pi/2$. This means that
the polar axis of the spherical polar coordinate system is chosen to be
perpendicular to the plane containing the origin and the arbitrarily
chosen points $(R,\Omega)$ and $(R,\Omega^\prime)$ on the PSS. Because
of the spherical symmetry. the result obtained will be valid for any
orientation of the axes of the spherical polar coordinate system.
When $\theta=\theta^\prime=\pi/2$, one finds that
$\partial\gamma/\partial\theta=\partial\gamma/\partial\theta^\prime=0$ and
$\partial^2\gamma/(\partial\theta\,\partial\theta^\prime)=-1/\sin\gamma$.
Hence,
\begin{equation}
\left( {{\partial^2S}\over{\partial\theta\,\partial\theta^\prime}} \right)
_{\theta=\theta^\prime=\pi/2}
=-{{1}\over{\sin\gamma}}\,{{\partial S}\over{\partial\gamma}}
\label{B4}
\end{equation}
The other second derivatives can be evaluated in the same way. We
obtain
\begin{eqnarray}
&\displaystyle\left( {{\partial^2S}\over{\partial\theta\,\partial\phi^\prime}}
\right)_{\theta=\theta^\prime=\pi/2}
=\left( {{\partial^2S}\over{\partial\phi\,\partial\theta^\prime}} \right)
_{\theta=\theta^\prime=\pi/2} =0 \nonumber\\
&\displaystyle\left( {{\partial^2S}\over{\partial\phi\,\partial\phi^\prime}}
\right)_{\theta=\theta^\prime=\pi/2}
=-{{\partial^2S}\over{\partial\gamma^2}}
\label{B5}
\end{eqnarray}
Hence, the determinant in (\ref{B1}) has the value given in equation
({\ref{eq4.5}).

\figure{Examples of trajectories in two dimensions which start from an
arbitrary point $\phi$ on the Poincar\'e circle and return
(in the same sense) to the Poincar\'e circle at the point $\phi^\prime$.
In each case there is a main trajectory together
with trajectories having slightly larger and slightly smaller angular
momenta. Fig.\ 1(a) was computed for the potential $V(r)\sim r^k$ with
$k=0.5$. Fig.\ 1(b) was computed for $V(r)\sim r^k$ with $k=3.5$.
\label{fig1}}

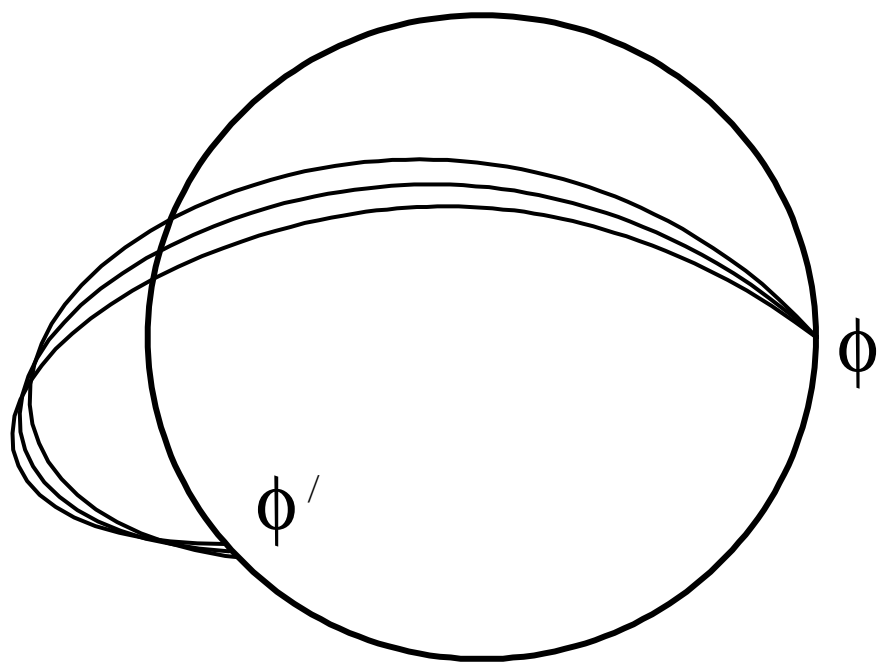
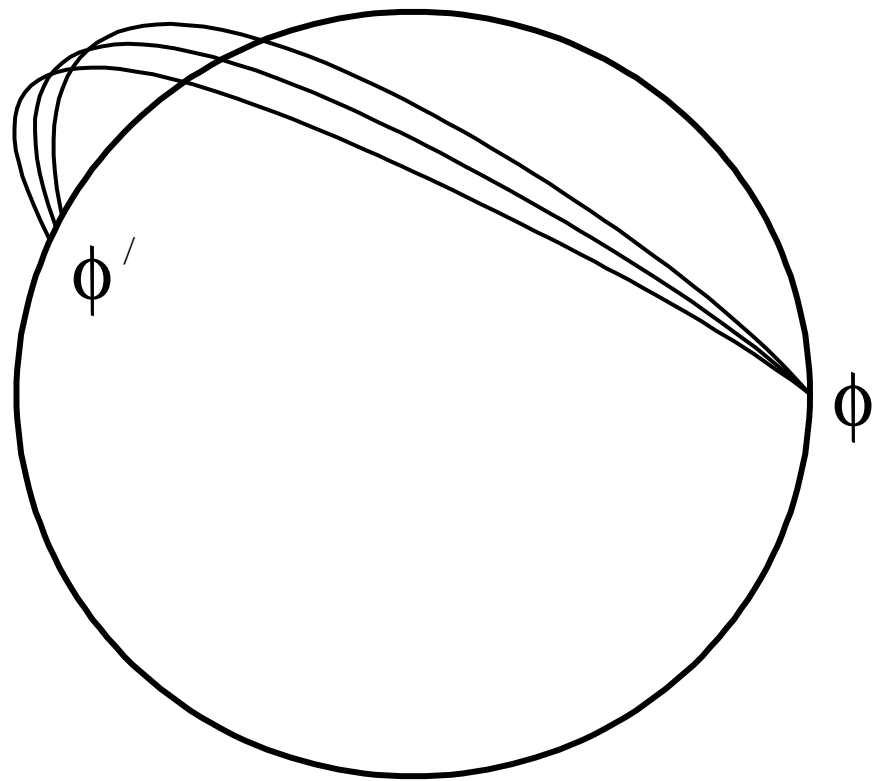
\end{document}